%% file: latexnb.tex
\documentclass[a4paper, review]{jds}      %

\jdsmonth{}                 %
\jdsyear{}                  %
\jdsvolume{xx}                  %
\jdsissue{xx}                   %
\jdsdoi{xx.xxxx/xxxxxxxxx}      %
\jdsreceived{}    %
\jdsaccepted{}      %

\usepackage{framed}
\usepackage[cache,nominted]{runcode}
\usepackage{tikz}
\usetikzlibrary{shapes.geometric}
\usetikzlibrary{positioning}

\title[Reproducible Science with \LaTeX]{Reproducible Science with \LaTeX}

\author{Haim Bar\thanks{Corresponding author. Email: haim.bar@uconn.edu}}
\author{HaiYing Wang}
\affil{Department of Statistics, University of Connecticut,
  Storrs, CT, 06269-4120, USA}

\begin{document}

\maketitle

\begin{abstract}
This paper proposes a procedure to execute external source codes from a \LaTeX\space document and include the calculation outputs in the resulting Portable Document Format (pdf) file automatically. It integrates programming tools into the \LaTeX\space writing tool to facilitate the production of reproducible research. In our proposed approach to a \LaTeX-based scientific notebook the user can easily invoke any programming language or a command-line program when compiling the \LaTeX\space document, while using their favorite \LaTeX\space editor in the writing process. The required \LaTeX\space setup, a new \proglang{Python} package, and the defined preamble are discussed in detail, and working examples using \proglang{R}, \proglang{Julia}, and \proglang{MatLab} to reproduce existing research are provided to illustrate the proposed procedure. We also demonstrate how to include system setting information in a paper by invoking shell scripts when compiling the document.\@
\end{abstract}


\section{Introduction}
\input{Introduction}

\section{Software setup}
We present the \LaTeX\space setup, server configuration, and some short commands specific to certain programming languages for ease of use.
The \textit{runcode} package can be used in two modes -- `batch' and `server'. When the \LaTeX\space document in compiled in the batch mode, pdflatex invokes the command-line statistical tool each time it executes a code segment and when that execution is complete, the statistical software exists. For example, with \proglang{R} batch-mode execution is done by \textit{runcode} like this:\\
\texttt{Rscript programName.R $>$ outputFile}\\
In the `server-mode' \textit{runcode} maintains a continuous bi-directional communication channel to the statistical software. The machinery enabling the server-mode is described in Section \ref{talk2stat}. In general, the batch-mode is simpler to use (and implement) since it does not require any additional packages, whereas server-mode requires the \proglang{Python} package \textit{talk2stat}. However, we recommend the server-mode since it maintains a `live session' which is more efficient -- both because the statistical software executable does not have to be invoked multiple times, and also because the continuous session makes all previous computations and environment variables readily available for subsequent operations, whereas in batch-mode it is necessary to  save and restore sessions between invocations for this purpose. 
Note that both modes can be used in the same \LaTeX\space document.

In the following it is assumed that the executables for the statistical software and for \proglang{Python} are in the user's path. If that is not the case or if the user wishes to specify a different path, this can be done by modifying the runcode.sty file, and adding the complete path to the executable file name (e.g. Rscript.)

\subsection[LaTeX setup]{\LaTeX\space setup}\label{setup}
\input{LatexSetup}

\subsection[talk2stat Python package]{\pkg{talk2stat} Python package}\label{talk2stat}

\input{talk2stat}

\subsection[Short commands for R, Julia, and MatLab]{Short commands for \proglang{R}, \proglang{Julia}, and \proglang{MatLab}}\label{alias}

\input{alias}

\section[An Example using R]{An Example using \proglang{R}}\label{exampleR}
\input{Rxample}

\section[Using Julia, MatLab, and R]{Using \proglang{Julia}, \proglang{MatLab}, and \proglang{R}}\label{exampleJuliaMatlab}
\input{JuliaMatlabR}

\section{Conclusion}
\label{sec:discussion}
To easily reproduce results of software-based scientific analyses we have developed a \LaTeX-based approach which automatically runs the code associated with the research and embeds its output when the document is compiled into a pdf file. Our approach is very general, in that any programming language or command-line tool can be used, while allowing \LaTeX\space users to continue to write their paper with their favorite text editor.

In this paper we explained in detail how to use our method, and described its important features, which include the ability to run external code, as well as to embed code directly in the tex file; the flexibility to choose the output format in the final document (inline, pure \LaTeX, or verbatim, as the case warrants); and how to use it efficiently, by using cached results when needed.

The method is based on the so-called `shell-escape' built-in functionality in the \LaTeX\space engine. In doing so, a user has to be mindful about the security implications, since shell-escape does not perform any code validation. However, this is a non-issue \textit{unless} the user who runs the \LaTeX\space compiler has a highly privileged account, \textit{and} if the source of the code is dubious and unverified by the user. This is an unlikely combination of circumstances, and can be easily avoided by using best practices. Furthermore, the `shell-escape' option is required by many widely used \LaTeX\space packages such as the \pkg{TikZ} and \pkg{PGF} packages for creating graphics, and the \pkg{minted} package we adopted in this paper for syntax highlighting.

Finally, we point out that the method presented in this paper is broadly applicable, and not limited to just compiling \LaTeX\space documents. For example, using the same infrastructure, namely the \pkg{talk2stat} package, one can build a web-based graphical user interface or dashboard to perform statistical analysis.
To demonstrate this capability, we have implemented a very simple web application where a user can perform a binary classification of a given dataset, using one of seven different classifiers (4 in \proglang{R} and 3 in \proglang{Julia}). The demo is available at \url{http://137.99.56.173:8081/}.
In this example, Node JS is used for the front-end, and connects to \proglang{R} and \proglang{Julia} in the same way that was demonstrated in this paper with pdflatex. The \pkg{shiny} package in  \proglang{R} \citep{shiny} offers a similar functionality, but our approach is more flexible, in that it allows the developers to use their preferred tools, both for the front-end (e.g., Node, React, etc.) and the back-end (\proglang{R}, \proglang{Julia}, \proglang{MatLab}, \proglang{Python}.) As such, it is possible to develop smart-phone apps which can perform heavy computational tasks, using server-side statistical software. 

\appendix\label{appendix}
\section*{Computational details}
The following script includes general system information, software versions (Mac), \proglang{R}, \proglang{MatLab}, and \proglang{Julia} versions, as well as package versions (\pkg{Matrix}, \pkg{ineq}, and \pkg{igraph}, which we used earlier.)

\showCode{bash}{Code/paperconfig.sh}
\runExtCode{sh}{Code/paperconfig.sh}{sysinfo}
\includeOutput{sysinfo}

\begin{center}
{\large\bf SUPPLEMENTARY MATERIAL}
\end{center}

\begin{description}

\item[Title:] Minimum working examples for Reproducible Science with \LaTeX. (pdf) %

The file supplement.pdf provides minimum working examples in \proglang{R} for the procedure proposed in the main paper.
The examples are also available on github, at\\ \url{https://github.uconn.edu/hyb13001/LaTeXwithCode}

\end{description}

\bibliographystyle{chicago}
\bibliography{refs}

\end{document}

%% file: Introduction.tex
Reproducing research results has always been key to good science, but in recent years the topic rose to the forefront in the scientific community and the media. Famously, in 2005, John Ioannidis wrote a paper titled `Why Most Published Research Findings Are False' \citep{Ioannidis2005} and ignited a lively discussion and prompted many ideas how to make things better.
The expectation is, and should be, that given a careful recipe which describes how an experiment was done or a data set has been analyzed, others will achieve the same results if they follow the same procedure.
In the context of statistical analysis this means that authors are required to provide the necessary data and programs, as well as to specify dependencies on external resources (e.g., packages, data repositories, etc.) and settings (including random seeds, for example.)

The notion of a `scientific notebook software' has evolved greatly over the last couple of decades. A scientific notebook makes it possible for computational and statistical results to be generated by the same tool which is used to write the paper, thus ensuring that results reported in the paper match the most up to date analysis performed by the authors. Proprietary software such as \proglang{MatLab} and \proglang{Mathematica} have built-in notebooks, and open-source languages like \proglang{R} and \proglang{Python} have similar implementations,
namely, the \pkg{Sweave} package \citep{leisch2002sweave}, the \pkg{knitr} package \citep{knitr1,knitr2,knitr3} and the \pkg{rmarkdown} package \citep{rmarkdown1,rmarkdown2} for \proglang{R},
and the \pkg{Pweave} package and \pkg{Jupyter} for \proglang{Python}. The aforementioned tools insert prose into programming code, use the programming language to generate tex files, and then generate pdf files.

In this paper we take a different approach and incorporate the code in a \LaTeX\space document, rather than inserting prose into a program. 
When the document is compiled, the code is executed and the results are embedded in the pdf file. 
Our approach has a couple of advantages over the aforementioned programming language-based 
notebooks. First, any programming language can be easily used. In fact, any command-line executable can be invoked when preparing the pdf file from the tex file. While some integrated development environments (IDEs) support multiple languages, using more than one in the same document is, at best, cumbersome.
Second, as far as text editing goes, programming language editors are less convenient than \LaTeX\space editing tools such as TeXworks, Kile, TeXshop, or Emacs, just to name a few. This is obviously a subjective preference, but in our experience the available \LaTeX\space editing tools have a lot of features that (currently) most programming language editors 
do not have. In typical manuscripts more space is ultimately used for prose than for the computational results, so it makes sense to use the tool which is most convenient for the majority of the writing process.
Furthermore, the programming language-based notebook 
approach requires learning the markdown syntax, which is not as rich as what \LaTeX\space already offers. We want to emphasize that we by no means claim that the proposed approach aims to replace a programming language editor. In the stage of developing code for analysis, the specific editors for the used languages are preferable because they often provide rich support for coding and debugging that a \LaTeX editor does not have. However, when the code is relative stable and the focus becomes writing a report or a paper, the proposed procedures is recommended.

Our proposed approach achieves the same purpose of reproducibility as the well-known notebooks, but it is a lot more flexible. It is in the spirit of Donald Knuth's literate programming \citep{Knuth1984} which considers computer programs `to be works of literature'.  It is also a realization of the concept of a compendium introduced by \cite{gentleman2007statistical}.
The method uses basic features in \LaTeX, and the key is to enable the `shell-escape' option when compiling a \LaTeX\space document. Our \LaTeX-driven approach is very simple to implement. Since `shell-escape' is part of the \LaTeX\space core functionality, only a handful of external packages have to be included in the preamble.

In Section \ref{setup} we introduce the components of our proposed method. In Sections \ref{exampleR} and \ref{exampleJuliaMatlab} we demonstrate the \LaTeX-based notebook approach using \proglang{R}, %
\proglang{Julia}, and \proglang{MatLab}.
Reproducibility requires authors to provide not only code and data, but also system configuration at the time of the compilation.
Each operating system has its own functions to report configuration settings, and in the Appendix we show an example on a Mac. This also serves as an example of how to invoke a shell script.

%% file: LatexSetup.tex
Our proposed approach requires to enable the `shell-escape' option when compiling a \LaTeX\space document. From the command line, it is done like this
\vspace{2mm}\\
\indent \verb|latex --shell-escape myFile.tex|
\vspace{2mm}\\
Most people use text editors that have \LaTeX\space compilation capability, which can be configured to enable this option. Here are examples using some commonly used text editors.
\begin{itemize}
\item With TeXnicCenter: under ``Build > Define Output Profile > Latex $\Rightarrow$ PDF'', insert the option \verb|-shell-escape| into ``Command line arguments to pass to the compiler''.
\item With TeXshop: under ``Preferences > Engine > pdfTeX > LaTeX Input Field'', change\\
\indent \verb|pdflatex --file-line-error --synctex=1|\\
to\\
\indent \verb|pdflatex --file-line-error --synctex=1 --shell-escape|.
\item With TeXStudio: under ``Options > Configure TeXStudio > Commands'',
change\\
\indent \verb|pdflatex -synctex=1 -interaction=nonstopmode %.tex|\\
into\\
\indent {\small\verb|pdflatex -synctex=1 -interaction=nonstopmode --shell-escape %.tex|}.
\item With TeXworks: in ``Preferences > Typesetting > pdflatex'', add \verb|--shell-escape| as an argument  and move it before the argument called `\$fullname'.
\end{itemize}
For other text editors, see their specific documentation to see how it is done.

It is also worth mentioning the popular web-based editor and collaboration tool for \LaTeX, \href{https://www.overleaf.com/}{Overleaf}. It is possible to use knitr on Overleaf, which allows authors to include \proglang{R} in their manuscripts. Because the `shell-escape' option is enabled by default on the Overleaf server, our method can also be used on Overleaf by simply installing the runcode.sty file in the project directory. Currently, both \proglang{R} and \proglang{Python} are installed on \href{https://www.overleaf.com/}{Overleaf}, but \proglang{Julia} and \proglang{MatLab} are not. Whenever a language other than \proglang{R} and \proglang{Python} becomes available on Overleaf, it will be immediately possible for Overleaf users of \textit{runcode} to take advantage of it.
Note that currently runcode can only be used in the batch-mode on OverLeaf, but we hope that in the future Overleaf will enable the server-mode functionality.

Our \LaTeX-driven approach requires a few external packages in the preamble. 
The packages \pkg{minted}, \pkg{fontenc}, and \pkg{textgreek}, are used in order to ensure that code and output are displayed correctly and aesthetically, even when they contain non-ASCII characters. Note that the package \pkg{minted} requires \proglang{Python} and the \proglang{Python} package \pkg{pygments} in order to run. Without \pkg{pygments}, one can use the package \pkg{fvextra} instead of  \pkg{minted}. However, \pkg{fvextra} does not provide syntax highlights\footnote{The reason we prefer to use minted and fvextra over the listings package is that the former support the extended Unicode character set. In addition, the minted package uses the pygmentize parsing library which is more extensive and expressive for programming languages compared with the keyword detection in listings.}. To embed code in \LaTeX\space source file directly, we use the \pkg{filecontents} package.
The \pkg{filecontents} package is only explicitly loaded if an older version of \LaTeX\space is used. It has been part of the \LaTeX\space engine since 2019. 
The packages \pkg{tcolorbox} and \pkg{xcolor} are used to enhance the output, and the \pkg{xifthen}, \pkg{xparse}, and \pkg{xstring} packages are used in the implementation of our new commands to make them clearer and more efficient. We put all the \LaTeX\space setups in a new package called \pkg{runcode}, and the style file runcode.sty can be loaded by \verb|\usepackage[options]{runcode}|, where possible options are: %
  \begin{itemize} \setlength\itemsep{-0.1em}
  \item `run': run source code by setting the global Boolean variable
  \texttt{runcode} to be true;
  \item `cache': use cached results by setting the global Boolean variable
  \texttt{runcode} to be false;
  \item `R': start server for \proglang{R} (this requires the \proglang{Python} \pkg{talk2stat} package);
  \item `julia'; start server for \proglang{Julia}  (this requires the \proglang{Python} \pkg{talk2stat} package);
  \item `matlab': start server for \proglang{MatLab}  (this requires the \proglang{Python} \pkg{talk2stat} package); and/or
  \item `nominted': use the \pkg{fvextra} package instead of the \pkg{minted} package to show code (this does not require the \proglang{Python} \pkg{pygments} package, but it does not provide syntax highlights).
  \end{itemize}

We have implemented four basic commands in the \pkg{runcode} package: %
\begin{itemize}
\item \verb|\runExtCode{Arg1}{Arg2}{Arg3}[Arg4]| runs an external
  code. It takes four arguments: \verb|Arg1| is the executable
  program, \verb|Arg2| is the source file name, \verb|Arg3| is the
  output file name (with an empty value, the counter
  \verb|codeOutput| is used), and \verb|Arg4| controls when to run the
  code. \verb|Arg4| is an optional argument with three possible
  values: if skipped or with empty value, the value of the global Boolean variable
  \texttt{runcode} is used; if the value is set to `run', the code
  will be executed; if set to `cache' (or anything else), use cached
  results (see more about the cache below).

\item \verb|\showCode{Arg1}{Arg2}[Arg3][Arg4]| shows the source code,
  using \pkg{minted} for a pretty layout or \pkg{fvextra} (if
  \proglang{Python} or \pkg{pygments} are not installed). It takes four
  arguments: \verb|Arg1| is the programming language, \verb|Arg2| is
  the source file name, \verb|Arg3| and \verb|Arg4| are the first and
  last line to show (optional - if skipped or with empty values, all the lines in the
  source file will be printed).
\item \verb|\includeOutput{Arg1}[Arg2]| is used to embed the output
  from executed code.  It takes two arguments: \verb|Arg1| is
  the output file name, and it needs to have the same value as that of \verb|Arg3| in \verb|\runExtCode|. If an empty value is given to \verb|Arg1|, the counter \verb|codeOutput| is used. \verb|Arg2| is optional and it controls the type output with a default value `vbox' (skipped or `vbox' = verbatim in a box, `tex' =
  pure latex, or `inline' = embed result in text). 
\item \verb|\inln{Arg1}{Arg2}[Arg3]| is designed for simple calculations; it runs one command (or a short batch) and displays the output within the text. \verb|Arg 1| is the executable program or programming language. \verb|Arg2| is the source code. \verb|Arg 3| is the output type  (if skipped or with empty value, the default type is `inline'; `vbox' = verbatim in a box).
\end{itemize}

Typically, at least in our experience, the writing of the text portion of a manuscript begins when the code is relatively stable, at which point most of the edits are applied to the text. In this situation, one may want to disable the execution of the embedded code and use the cached results from the last compilation, so that subsequent \LaTeX\space compilations are faster. We define a Boolean variable \texttt{runcode} to allow this functionality, and one can disable the execution of the embedded code by declaring the `cache' option when loading the \pkg{runcode} package (this sets \texttt{runcode} to false). When code execution is disabled, output from the last time the code was run is used when compiling the tex file. As discussed above, one can override the global setting of \texttt{runcode} in order to enable or disable the execution of selected chunks of code.
All calculation results that can be included in \LaTeX\space output are stored in the folder ``tmp'' in the directory of the working \LaTeX\space file.

In the following sections we demonstrate the \LaTeX-based notebook approach using \proglang{R}, %
\proglang{Julia}, \proglang{MatLab}/\proglang{Octave}, and shell scripts.

%% file: talk2stat.tex
The functionality of a new \proglang{Python} package \pkg{talk2stat} is described in the following figure. In this example, we show a situation in which \proglang{Julia} is used as the statistical engine.\\
\begin{center}
\begin{tikzpicture}
  \node[draw, minimum size=1cm]  (talk2stat) {\Large\pkg{talk2stat}};
  \node[minimum size=1cm, left = 2cm of talk2stat] (pdflatex) {\Large pdflatex};
  \node [cylinder, shape border rotate=180, draw, minimum height=25mm, minimum width=5mm,  right = 7mm of  talk2stat] (pipe) {bi-directional pipe};
  \node[right = 7mm of pipe] (julia) {\Large \proglang{Julia}};
  \node[below = 15mm of talk2stat] (fs) {\Large File system};
  \draw [<->,line width=0.5mm,blue,dashed] (pdflatex) edge node[above] {Socket} (talk2stat);
  \draw [<-,line width=0.5mm,orange] (talk2stat) edge (pipe);
  \draw [->,line width=0.5mm,orange] (pipe) edge (julia);
   \draw [<->,line width=0.5mm,red, dotted] (julia) edge (fs);
   \draw [<->,line width=0.5mm,red, dotted] (pdflatex) edge (fs);
   \draw [<->,line width=0.5mm,red, dotted] (talk2stat) edge (fs);
\end{tikzpicture}
   \end{center}
   
When a tex file is being compiled, the \pkg{runcode} package first checks if the \pkg{talk2stat} server is running, and if not, it starts it. The server listens to a user-defined port number, and expects connections via a socket interface (using the Internet Protocol v4 addresses) which is represented by the dashed blue line. The default port numbers in the \pkg{runcode} package for \proglang{R}, \proglang{Julia}, and \proglang{MatLab} are 65432, 65431, and 65430, respectively.

During the compilation of the tex file, the following steps occur each time an embedded code is encountered (using \verb|\runExtCode| or \verb|\inln|).
\begin{enumerate}
 \item \pkg{runcode} reads the code, which can be specified either a file name containing the code, or a short  in-line code segment.
 \item The code or input file name is sent to \pkg{talk2stat}'s client function, which uses the socket interface to accept requests. 
 \item \pkg{talk2stat}'s server function sends the code to the statistical engine via a `bi-directional pipe'\footnote{A bi-directional pipe allows two programs to communicate in a synchronous fashion by sending one program's output to the other's input}. This is represented by the solid orange line in the figure.
 \item The statistical engine runs the code, and prints the output to the standard output via the pipe. The code may contain instructions to print output to the file system (e.g., figures.)
 \item \pkg{talk2stat}'s server function writes the output from the statistical engine to the output file which is specified in  \verb|\runExtCode| or \verb|\inln|.
 \item pdflatex recognizes that the operation invoked by \verb|\runExtCode| or \verb|\inln| is complete, embeds the output in the pdf file (via \verb|\include| or \verb|\includegraphics|) and moves on to the next line in the tex file. 
\end{enumerate}
All three components in the system, namely, pdflatex, \textit{talk2stat}, and the statistical engine, may use the file system to read data or scripts or to write their output (text, tables, figures, etc.). This is represented by the dotted red lines in the diagram.

%% file: alias.tex
When writing in \LaTeX, many people define some custom shortcuts to facilitate the writing. For the same purpose, we define some short commands specifically for \proglang{R}, \proglang{Julia}, and \proglang{MatLab}. These short commands are based on the \verb|\runExtCode| and \verb|\inln| commands presented in Section \ref{setup}, and they are customizable. In this paper, the main purpose of defining the short commands is for illustration, and a user can easily use the similar style to define short commands for other programming languages. All the short commands use the running \pkg{talk2stat} server to execute source codes if the executable program is not specified. 

The short commands defined for \proglang{R}, \proglang{Julia}, and \proglang{MatLab} are listed below.
\begin{itemize}
\item \proglang{R}
  \begin{itemize}
  \item \verb|\runR[Arg1]{Arg2}{Arg3}[Arg4]| runs an external \proglang{R} code file. \verb|Arg1| is optional and uses \pkg{talk2stat}'s  \proglang{R} server by default (rather than invoking \texttt{Rscript}). \verb|Arg2|, \verb|Arg3|, and \verb|Arg4| have the same effects as those of \verb|\runExtCode| in Section \ref{setup}.
  \item \verb|\inlnR[Arg1]{Arg2}[Arg3]| runs \proglang{R} source code (\verb|Arg2|) and displays the output in line. \verb|Arg1| is optional and uses the \proglang{R} server by default. \verb|Arg2| is the \proglang{R} source code to run. If the \proglang{R} source code is wrapped between \verb|```| on both sides (as in the markdown grammar), then it will be implemented directly; otherwise the code will be written to a file on the disk and then be called. \verb|Arg3| has the same effect as that in the definition of \verb|\inln| in Section \ref{setup}.
\end{itemize}

\item \proglang{Julia}
  \begin{itemize}
  \item \verb|\runJulia[Arg1]{Arg2}{Arg3}[Arg4]| runs an external \proglang{Julia} code file. \verb|Arg1| is optional and uses \pkg{talk2stat}'s \proglang{Julia} server by default. \verb|Arg2|, \verb|Arg3|, and \verb|Arg4| have the same effects as those of \verb|\runR|.
  \item \verb|\inlnJulia[Arg1]{Arg2}[Arg3]| runs \proglang{Julia} source code and displays the output in line. \verb|Arg1| is optional and uses \pkg{talk2stat}'s  \proglang{Julia} server by default. \verb|Arg2| and \verb|Arg3| have the same effects as those of \verb|\inlnR|.
\end{itemize}

\item \proglang{MatLab}
Similarly, we define
  \begin{itemize}
  \item \verb|\runMatLab[Arg1]{Arg2}{Arg3}[Arg4]|%
  \item \verb|\inlnMatLab[Arg1]{Arg2}[Arg3]|%
\end{itemize}
\end{itemize}

%% file: Rxample.tex
We demonstrate our approach using data and code provided with the paper
\textit{Coauthorship and citation networks for statisticians}
\citep{ji2016,jin2015}.
  In their famous paper, Pengsheng Ji and Jiashun Jin analyze trends and patterns in statistical research, as well as author productivity and centrality, and community structures. In order to do that, they look at papers published in four highly ranked journals in statistics (Annals of Statistics, Biometrika, JASA and JRSS-B) from 2003 to the middle of 2012.
After careful cleaning and resolving author names in citations, their dataset contains  3,248 papers and 3,607 authors.
They perform many analyses, of which we use only a few to show our approach. %

Listing \ref{readData} shows the code used by \cite{ji2016} to read in the data and compute the adjacency matrix. 
We have changed the code slightly to make it more efficient, by coercing  \texttt{authorPaperBiadj} to be a sparse matrix (using the \pkg{Matrix} package \citep{Matrix}; see line 7).
To include this listing, we put the following in the tex file:
\begin{verbatim}
\begin{codelisting}{Reading and preparing the coauthorship data
                              \citep{ji2016,jin2015}}
\showCode{R}{Code/JiJin2016.R}
\end{codelisting}
\end{verbatim}

\begin{codelisting}{Reading and preparing the coauthorship data \citep{ji2016,jin2015}}%
\showCode{R}{Code/JiJin2016.R}
\label{readData}
\end{codelisting}

If we only want to show certain lines, we can use the optional arguments. For example \verb|\showCode{R}{Code/JiJin2016.R}[17][19]|,  \verb|\showCode{R}{Code/JiJin2016.R}[17][]|, or \verb|\showCode{R}{Code/JiJin2016.R}[17]| only shows the lines 17-19 as below.

\showCode{R}{Code/JiJin2016.R}[17]

Note that using \verb|\showCode| does not cause the code to be executed. To execute it we use \verb|\runR|.
Also note that the code in Listing \ref{readData} does not produce any output. %
To run the code using \pkg{talk2stat}'s server, we put the following in the tex file:
\begin{verbatim}
\runR{Code/JiJin2016.R}{initprog}
\end{verbatim}
Note that to use the server, a user has to use the \verb|R| option when loading the \pkg{runcode} package, i.e., to use \verb|\usepackage[R]{runcode}|. All the codes are set to run by default globally. To disable code execution globally, use the \verb|cache| option, i.e., use \verb|\usepackage[cache, R]{runcode}|. We can also force the code to run or to use the cache by using the optional \verb|Arg4|. For example, we can use \verb|\runR{Code/JiJin2016.R}{initprog}[run]| to force the code to run, and use \verb|\runR{Code/JiJin2016.R}{initprog}[cache]| to prohibit the code from running. 
\runR{Code/JiJin2016.R}{initprog}%

If we want to start a new \proglang{R} session each time a code segment is executed instead of using the \proglang{R} server, we use
\begin{verbatim}
\runR[Rscript --save --restore]{Code/JiJin2016.R}{initprog}
\end{verbatim}
Here, the \verb|--save| option saves all variables created in this program so that they can be used by future program, and the \verb|--restore| option restores all variables created in the previous session. If we use \pkg{talk2stat}'s server to communicate with a statistical software (\proglang{R}, in this section) there is no need to save any variable or environment setting, since the  is \proglang{R} session  runs continuously and all variable are accessible at all time. Thus, the server option is recommended as it is more efficient than the `batch mode', in which \texttt{Rscript} is invoked for each code segment. We will use the server option by default for the rest of the paper except for the appendix, where we show how to embed system configuration in the paper. 

If we want to put \proglang{R} code in the tex file rather than in an external file, we do it by using the \textit{filecontents*} environment. %
For example, %
we can get a table of the number of papers by year by writing the following:
\begin{verbatim}

\runR{tmp/temp00.R}{paperYear}
\includeOutput{paperYear}
\end{verbatim}

\runR{tmp/temp00.R}{paperYear}

This produces the following:
\includeOutput{paperYear}

Note that the second mandatory argument in \verb|\runR| matches the mandatory argument in \verb|\includeOutput|, and it is the name of the file under the tmp folder which is used to save the output from the executable (in this case, an \proglang{R} program called tmp/temp00.R). The default value for the optional argument of \verb|\includeOutput| is \verb|vbox|, so \verb|\includeOutput{paperYear}| is equivalent to \verb|\includeOutput{paperYear}[vbox]|. The \verb|vbox| option is used to print the output verbatim (no \LaTeX \space formatting) and enclose it in a box.
The same result can also be obtained by the \verb|\inlnR| command as \verb|\inlnR{```table(paperYear)```}[vbox]|.

The \verb|\inlnR| command can be used when the output should be embedded in the text.
For example, in their paper, Ji and Jin report the Gini coefficient based on the paper-adjacency matrix, which they compute using the \pkg{ineq} package \citep{ineq}.
To execute their code and include its output in-line, we put the following in the tex document:
\begin{verbatim}
The Gini coefficient is \inlnR{```cat(format(ineq(rowSums(paperCitAdj)),
digits=2))```}, suggesting that the in-degree is highly dispersed.
\end{verbatim}

This produces the following: 
`The Gini coefficient is \inlnR{```cat(format(ineq(rowSums(paperCitAdj)),digits=2))```}, suggesting that 
the in-degree is highly dispersed.'

The \verb|\inlnR| command is designed to work for very simple calculations, and currently it does not allow \verb|"| sign in the code.

To demonstrate other output options, we use the code provided by  \cite{ji2016} to produce their Table 2 and Figure 4. For table 2, the code is saved in a file called Code/JiJin2016Table2.R, which requires
the \pkg{kableExtra} library to format the output as a \LaTeX \space table, and the package \pkg{igraph} \citep{igraph} to calculate the closeness and betweenness metrics. Since the code produces a \LaTeX-formatted table, we can enclose it in a \textit{table}  environment and achieve the same layout as in \cite{ji2016} (Table 1.)

\begin{verbatim}
\runR{Code/JiJin2016Table2.R}{table2}
\includeOutput{table2}[tex]
\end{verbatim}

\runR{Code/JiJin2016Table2.R}{table2}
\includeOutput{table2}[tex]

The code to produce Figure 4 in \cite{ji2016} is saved in Code/JiJin2016Plot4.R. The program creates two pdf files called tmp/LC-citations.pdf and tmp/propCitation.pdf. Using these pdf files and the \textit{figure} environment in \LaTeX\space we can achieve the same layout as in \cite{ji2016}, in Figure \ref{jj2016figure4} below.

\begin{verbatim}
\runR{Code/JiJin2016Plot4.R}{}
\begin{figure}[b!]
\begin{minipage}[b]{0.45\linewidth}
\centering
\includegraphics[width=\textwidth]{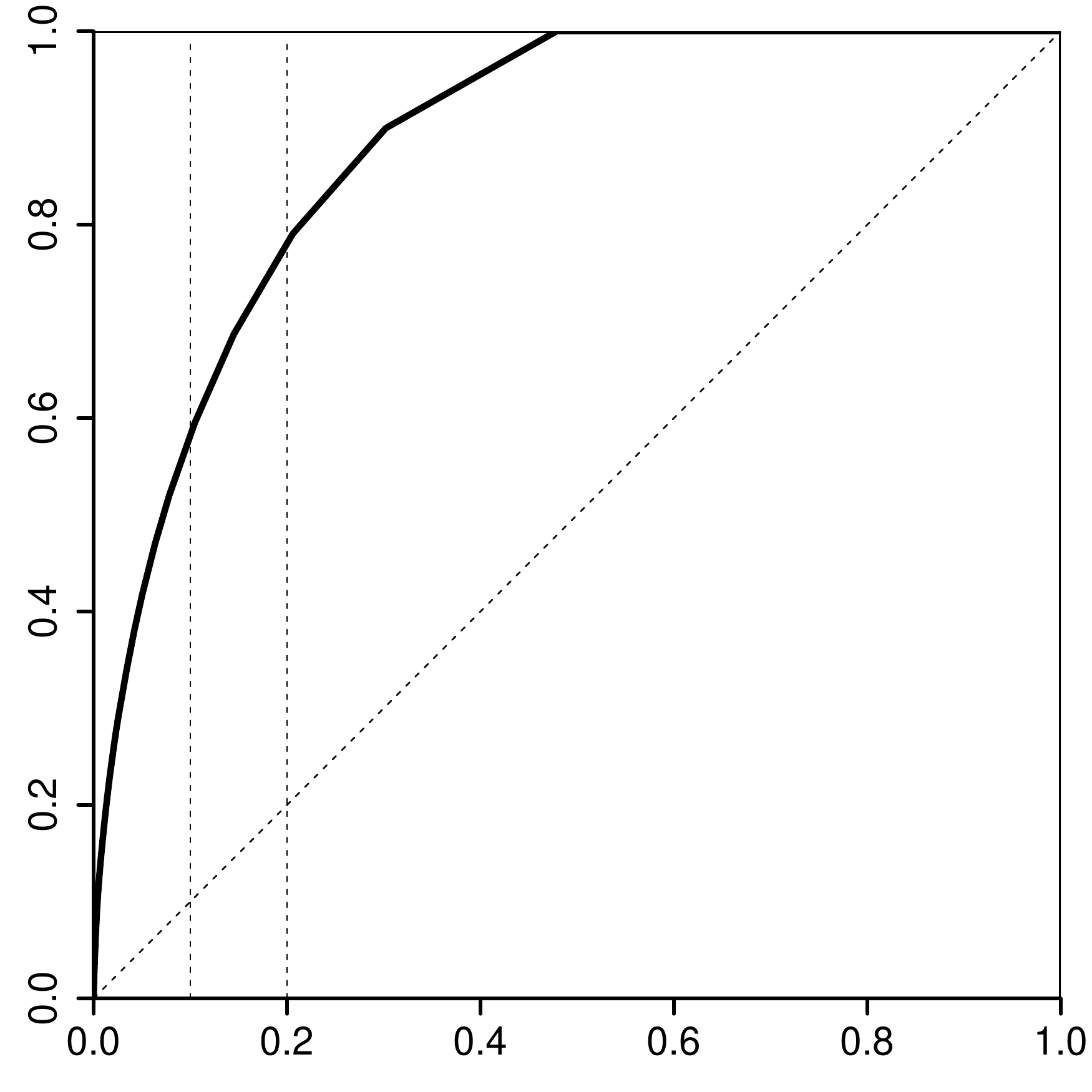}
\end{minipage}
\hspace{0.5cm}
\begin{minipage}[b]{0.45\linewidth}
\centering
\includegraphics[width=\textwidth]{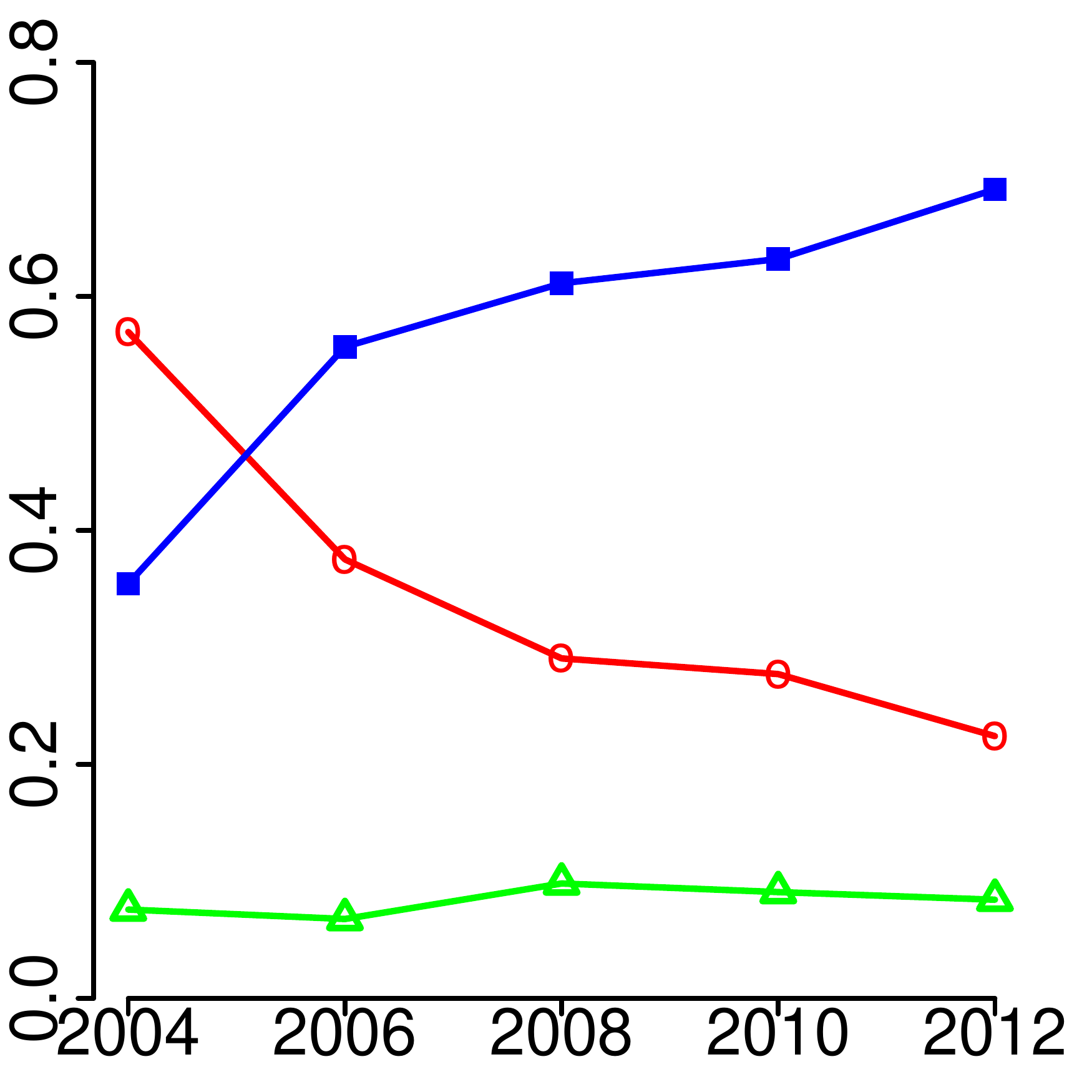}
\end{minipage}
\caption{Reproducing Figure 4 from \cite{ji2016}}\label{jj2016figure4}
\end{figure}
\end{verbatim}

\runR{Code/JiJin2016Plot4.R}{}
\begin{figure}[b!]
\begin{minipage}[b]{0.45\linewidth}
\centering
\includegraphics[width=\textwidth]{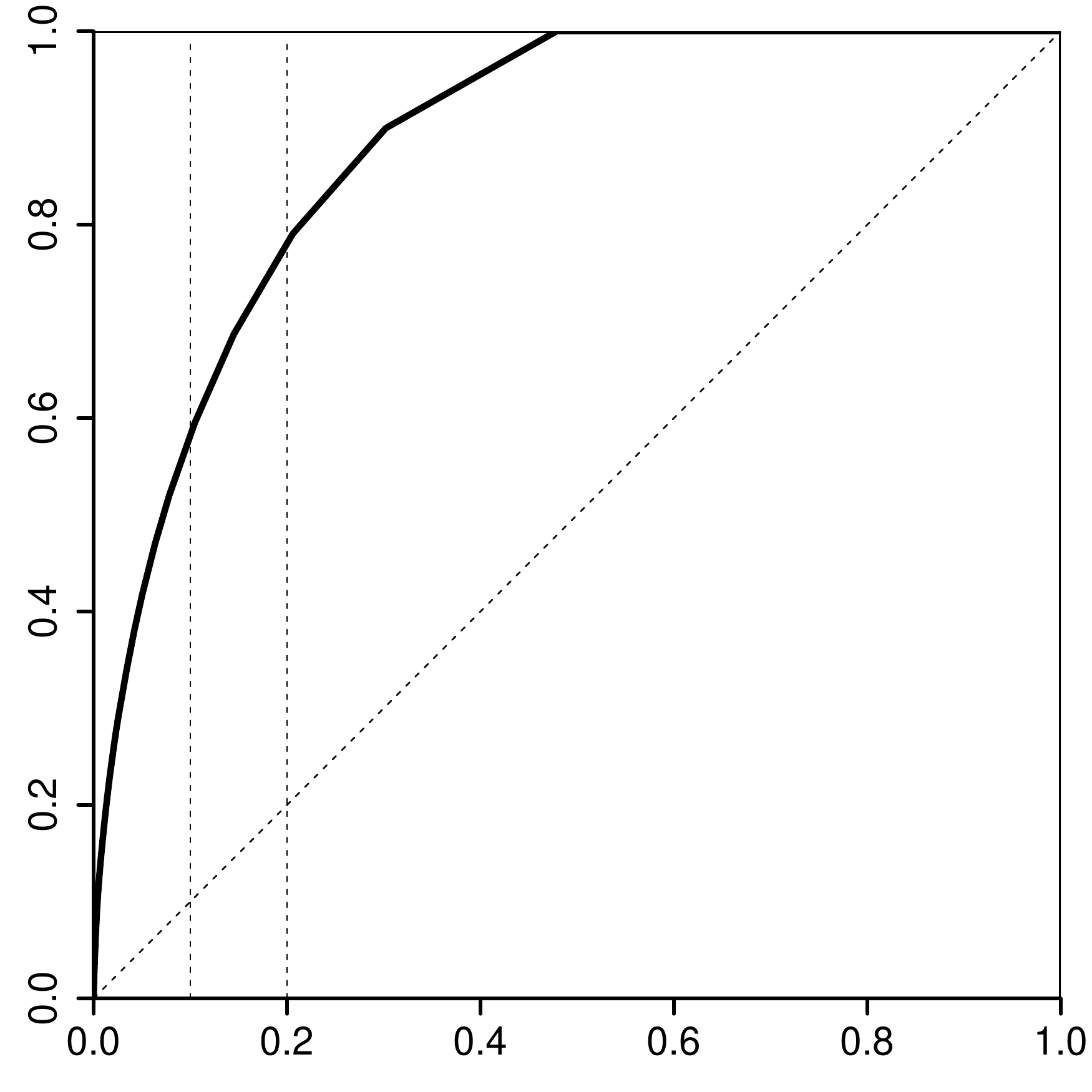}
\end{minipage}
\hspace{0.5cm}
\begin{minipage}[b]{0.45\linewidth}
\centering
\includegraphics[width=\textwidth]{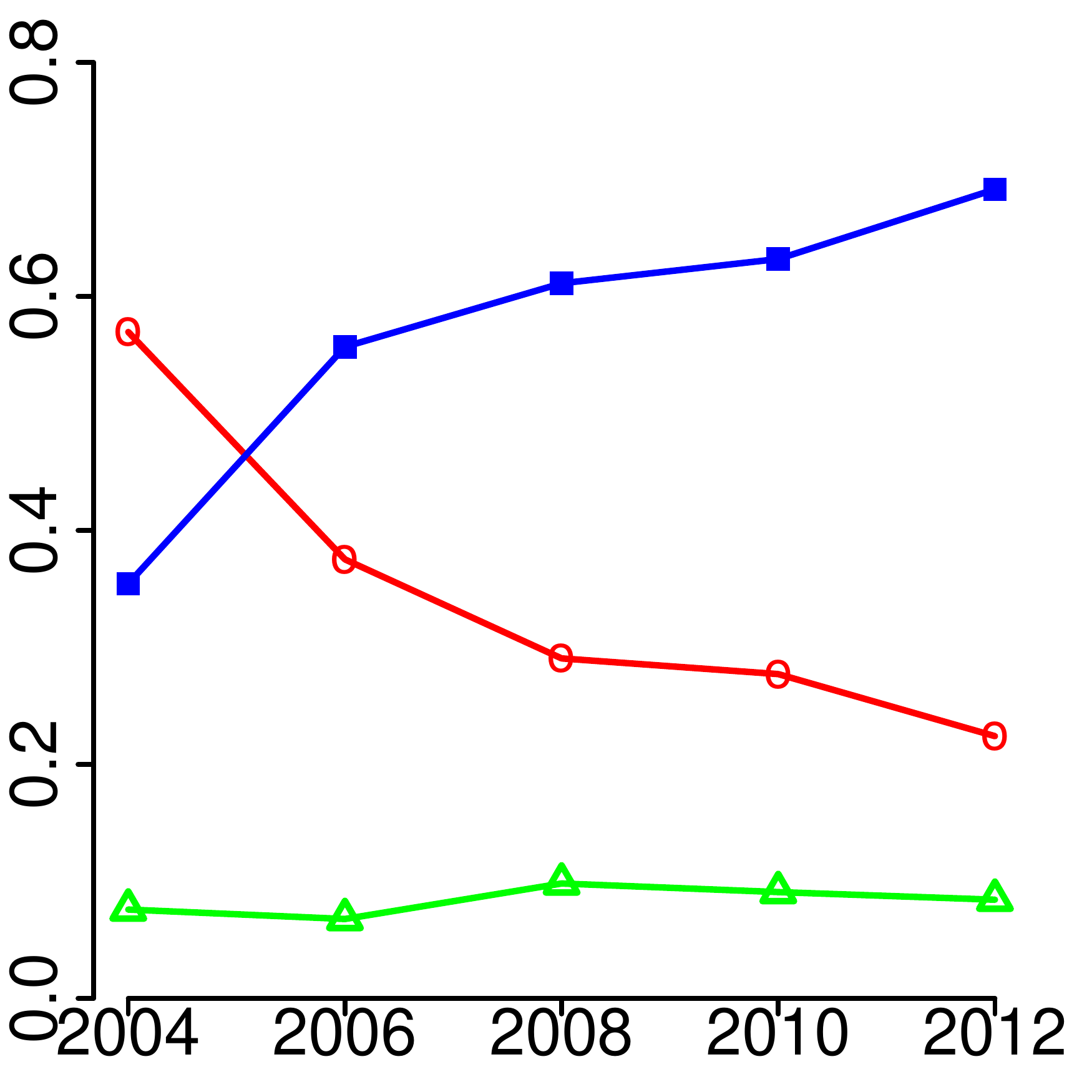}
\end{minipage}
\caption{Reproducing Figure 4 from \cite{ji2016}}\label{jj2016figure4}
\end{figure}

%% file: JuliaMatlabR.tex
It is common to use multiple programming languages in one project. For example, \cite{ji2016} use \proglang{R} and \proglang{MatLab} to generate Figure 6 in that paper. The proposed approach is particularly suitable for this scenario as one can incorporate all the external codes for multiple programming languages in \LaTeX. Here, as an illustration, we 1) use \proglang{Julia} to process the data and perform matrix operations; 2) use \proglang{R} and \proglang{MatLab} functions provided by the authors of \cite{ji2016} to find author clusters; and 3) use \proglang{R} to generate the figure. Since we use three different programming languages in this example, we use `R', `julia', and `matlab' options when loading the \pkg{runcode} package, i.e., use \verb|\usepackage[R,julia,matlab]{runcode}|.

To run the \proglang{Julia} code in file ``Code/JiJin2016Julia.jl'', we put the following in the tex file:
\begin{verbatim}
\runJulia{Code/JiJin2016Julia.jl}{initjulia}
\end{verbatim}
\runJulia{Code/JiJin2016Julia.jl}{initjulia}

The \proglang{Julia} code processes the data and find the 15 authors' names for Figure 6 of \cite{ji2016}. The 15 authors' names are stores in a variable called \verb|top15|. We can use the \verb|\inlnJulia| option to reproduce the second paragraph in Section 4.2 of \cite{ji2016} by writing the following in the tex file:
\begin{verbatim}
The giant component (236 nodes) is seen to be the ``High-Dimensional
Data Analysis [Coauthorship (A)]'' community (HDDA-Coau-A), including 
(sorted descendingly by the degree) \includeOutput{initjulia}[inline],
 etc. It seems that the giant component has substructures.
\end{verbatim}
This produces the following:\\
`The giant component (236 nodes) is seen to be the ``High-Dimensional
Data Analysis [Coauthorship (A)]'' community (HDDA-Coau-A), including 
(sorted descendingly by the degree) \inlnJulia{```print(top15)```}, %
 etc. It seems that the giant component has substructures.'

Now we run the \proglang{MatLab} code provided in \cite{ji2016}. The main file is ``MatlabCode.m'', and we run it from \LaTeX\ by including the following in the \LaTeX\ source file:
\begin{verbatim}
\runMatlab{matlabinterim}{}
\end{verbatim}
\runMatlab{matlabinterim}{}
We do not need to include any direct results from the \proglang{MatLab} code in the paper. Now we run the \proglang{R} code in ``JiJin2016Plot6.R'' to create Figure 6 of \cite{ji2016} by
\begin{verbatim}
\runR{Code/JiJin2016Plot6.R}{JiJin2016fig6}
\end{verbatim}
\runR{Code/JiJin2016Plot6.R}{JiJin2016fig6}

\begin{figure}[ht]
\centering
\includegraphics[width=0.8\textwidth,page=1]{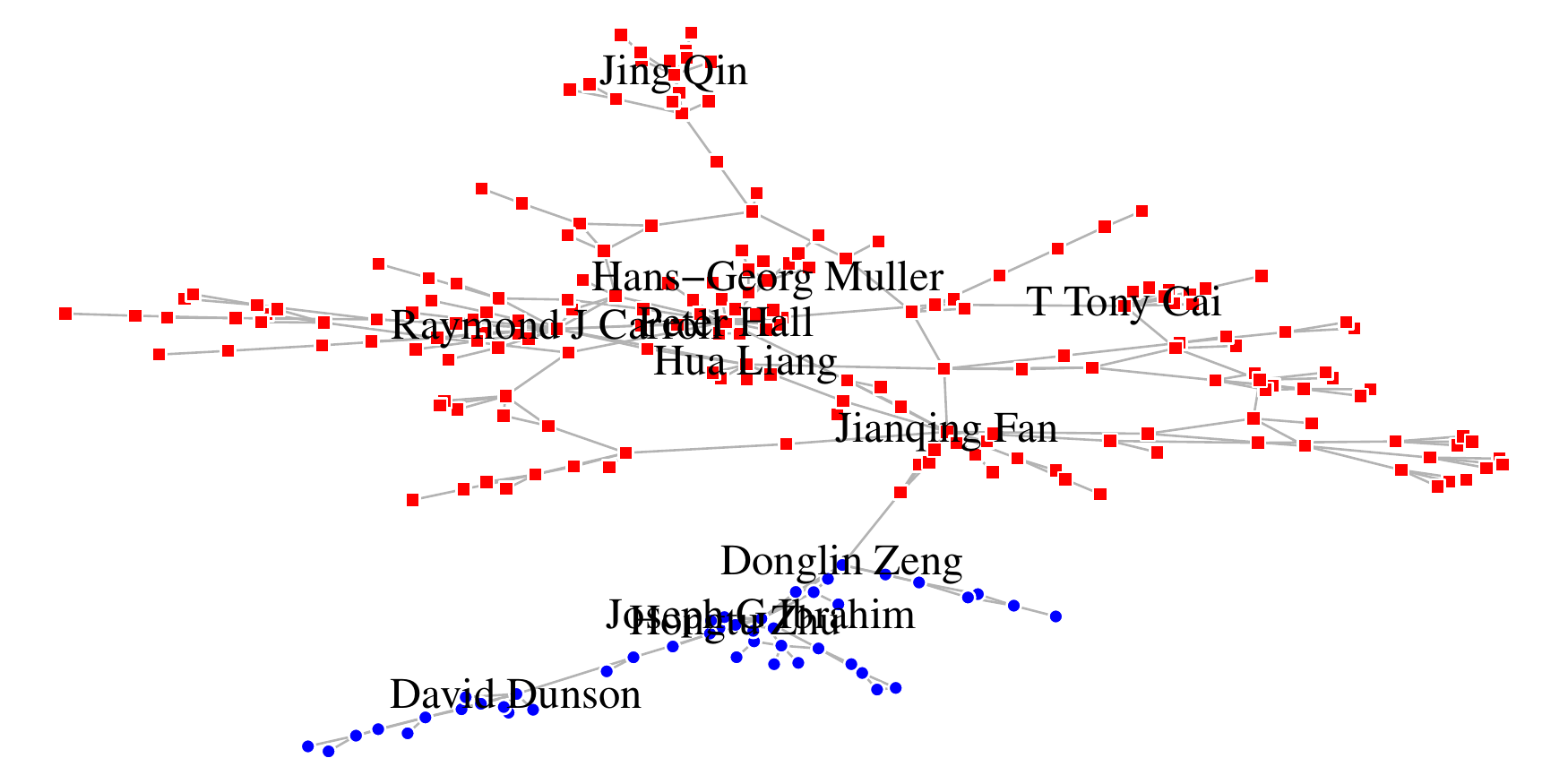}
\includegraphics[width=0.8\textwidth,page=2]{tmp/JiJin2016fig6.pdf}
\caption{Reproducing Figure 6 from Ji and Jin, 2016}\label{jj2016figure6}
\end{figure}

The resulting plots are shown in Figure~\ref{jj2016figure6}. Due to the updated \texttt{igraph} package, the resulting figures are not identical to Figure 6 of \cite{ji2016}, but the clusters and connections are the same.

%% file: latexnb.bbl
\begin{thebibliography}{}

\bibitem[\protect\citeauthoryear{Allaire, Xie, McPherson, Luraschi, Ushey,
  Atkins, Wickham, Cheng, Chang, and Iannone}{Allaire
  et~al.}{2020}]{rmarkdown1}
Allaire, J., Y.~Xie, J.~McPherson, J.~Luraschi, K.~Ushey, A.~Atkins,
  H.~Wickham, J.~Cheng, W.~Chang, and R.~Iannone (2020).
\newblock {\em rmarkdown: Dynamic Documents for R}.
\newblock {R} package version 2.1.

\bibitem[\protect\citeauthoryear{Bates and Maechler}{Bates and
  Maechler}{2019}]{Matrix}
Bates, D. and M.~Maechler (2019).
\newblock {\em Matrix: Sparse and Dense Matrix Classes and Methods}.
\newblock {R} package version 1.2-16.

\bibitem[\protect\citeauthoryear{Chang, Cheng, Allaire, Xie, and
  McPherson}{Chang et~al.}{2020}]{shiny}
Chang, W., J.~Cheng, J.~Allaire, Y.~Xie, and J.~McPherson (2020).
\newblock {\em shiny: Web Application Framework for R}.
\newblock R package version 1.5.0.

\bibitem[\protect\citeauthoryear{Csardi and Nepusz}{Csardi and
  Nepusz}{2006}]{igraph}
Csardi, G. and T.~Nepusz (2006).
\newblock The igraph software package for complex network research.
\newblock {\em InterJournal\/}~{\em Complex Systems}, 1695.

\bibitem[\protect\citeauthoryear{Gentleman and Temple~Lang}{Gentleman and
  Temple~Lang}{2007}]{gentleman2007statistical}
Gentleman, R. and D.~Temple~Lang (2007).
\newblock Statistical analyses and reproducible research.
\newblock {\em Journal of Computational and Graphical Statistics\/}~{\em
  16\/}(1), 1--23.

\bibitem[\protect\citeauthoryear{Ioannidis}{Ioannidis}{2005}]{Ioannidis2005}
Ioannidis, J. P.~A. (2005, 08).
\newblock Why most published research findings are false.
\newblock {\em PLOS Medicine\/}~{\em 2\/}(8).

\bibitem[\protect\citeauthoryear{Ji and Jin}{Ji and Jin}{2016}]{ji2016}
Ji, P. and J.~Jin (2016, 12).
\newblock Coauthorship and citation networks for statisticians.
\newblock {\em Ann. Appl. Stat.\/}~{\em 10\/}(4), 1779--1812.

\bibitem[\protect\citeauthoryear{Jin}{Jin}{2015}]{jin2015}
Jin, J. (2015).
\newblock Fast community detection by score.
\newblock {\em The Annals of Statistics\/}~{\em 43\/}(1), 57--89.

\bibitem[\protect\citeauthoryear{Knuth}{Knuth}{1984}]{Knuth1984}
Knuth, D.~E. (1984).
\newblock Literate programming.
\newblock {\em The Computer Journal\/}~{\em 27\/}(2), 97--111.

\bibitem[\protect\citeauthoryear{Leisch}{Leisch}{2002}]{leisch2002sweave}
Leisch, F. (2002).
\newblock Sweave: Dynamic generation of statistical reports using literate data
  analysis.
\newblock In {\em Compstat}, pp.\  575--580. Springer.

\bibitem[\protect\citeauthoryear{Xie}{Xie}{2014}]{knitr3}
Xie, Y. (2014).
\newblock knitr: A comprehensive tool for reproducible research in {R}.
\newblock In V.~Stodden, F.~Leisch, and R.~D. Peng (Eds.), {\em Implementing
  Reproducible Computational Research}. Chapman and Hall/CRC.
\newblock {ISBN} 978-1466561595.

\bibitem[\protect\citeauthoryear{Xie}{Xie}{2015}]{knitr2}
Xie, Y. (2015).
\newblock {\em Dynamic Documents with {R} and knitr\/} (2nd ed.).
\newblock Boca Raton, Florida: Chapman and Hall/CRC.
\newblock {ISBN} 978-1498716963.

\bibitem[\protect\citeauthoryear{Xie}{Xie}{2020}]{knitr1}
Xie, Y. (2020).
\newblock {\em knitr: A General-Purpose Package for Dynamic Report Generation
  in R}.
\newblock {R} package version 1.28.

\bibitem[\protect\citeauthoryear{Xie, Allaire, and Grolemund}{Xie
  et~al.}{2018}]{rmarkdown2}
Xie, Y., J.~Allaire, and G.~Grolemund (2018).
\newblock {\em R Markdown: The Definitive Guide}.
\newblock Boca Raton, Florida: Chapman and Hall/CRC.
\newblock {ISBN} 9781138359338.

\bibitem[\protect\citeauthoryear{Zeileis}{Zeileis}{2014}]{ineq}
Zeileis, A. (2014).
\newblock {\em ineq: Measuring Inequality, Concentration, and Poverty}.
\newblock {R} package version 0.2-13.

\end{thebibliography}
